\documentclass{article}
\usepackage{epsf}
\newcommand{\be}{\begin{equation}}
\newcommand{\ee}{\end{equation}}
\newcommand{\ba}{\begin{array}{c}}
\newcommand{\ea}{\end{array}}
\newcommand{\bqa}{\begin{eqnarray}}
\newcommand{\eqa}{\end{eqnarray}}

\begin{document}

\markboth{Z. X. Sun, L. Y. Xiao, Z. G. Xiao, H. Q. Zheng} {Model
dependent analyses on the $N_c$ dependence of the  $\sigma$ pole
trajectory}

%
%

\title{Model dependent analyses on the $N_c$ dependence of the  $\sigma$ pole trajectory\footnote{Talk
presented
 at
\textit{Conference on Non-Perturbative Quantum Field Theory:
Lattice and Beyond}, Guangzhou, China, Dec.16--18, 2004. }}

\author{Z. X. Sun, L. Y. Xiao, Z. G. Xiao\footnote{Email: xiaozhiguang@pku.edu.cn} and H. Q. Zheng
\vspace{0.2cm}\\
Department of Physics, Peking University,
Beijing, 100871, China\\
}

\maketitle


\begin{abstract}
We explore the nature  of the$\sigma$  or $f_0(600)$ meson using
large $N_c$ technique, assuming that the $\sigma$ pole dominates
low energy physics in the IJ=00 channel. We trace the $N_c$
dependence of the $\sigma$ pole position using [1,1] Pad\'e
approximation of $SU(3)$ chiral amplitude, [1,2] Pad\'e
approximation of $SU(2)$ chiral amplitudes,  and the PKU
parametrization form.  We find that in all 3 cases the $\sigma$
pole moves to the real axis of the complex $s$-plane.  Finally we
construct a pedagogical example using the $SU(2)$ Pad\'e amplitude
to show that when the analyticity property of the scattering
amplitude is violated, it is possible that the $\sigma$ pole
resides on the complex $s$ plane when $N_c$ is large.

\vspace{.3cm}

\vspace{.3cm}
\end{abstract}

\section{Introduction}

The concept of the spontaneous breaking of chiral symmetry plays a
very important role in understanding the strong interaction
dynamics. The existence of a broad pole structure in the IJ=00
channel $\pi\pi$ scattering~\cite{PDG04} is also very important in
understanding the mechanism of the spontaneous  breaking of chiral
symmetry. It has been demonstrated that the $\sigma$ pole is
necessary even in the theory with non-linear realization of chiral
symmetry,~\cite{XZ00}  in order to explain the steady rise of the
scattering phase shift in the IJ=00 channel. Efforts have been
made in determining the pole position of such a broad resonance,
which is a highly nontrivial task. Experimental efforts have been
made in determining the pole location.~\cite{exp_sigma} An
extensive analysis based on the use of Roy equations determined
the pole mass and width to be $M=470\pm 30$MeV and $\Gamma=590\pm
40$MeV, respectively.\cite{CGL01} A very similar result on the
pole location
 with $M=470\pm 50$MeV and $\Gamma=570\pm50$MeV is found by a
quite different approach.~\cite{zhou04} It is emphasized that
crossing symmetry places an important role in the determination of
the $\sigma$ pole.~\cite{zhou04}

 Despite of the successes in the determination
of the $\sigma$ pole, little is known about the nature of such a
pole. For example, it is not understood well whether such a pole
has anything to do with the `$\sigma$' in the linear $\sigma$
model or not. Various kinds of speculations and investigations can
be found in the literature on such an issue, but in our
understanding, no definite, sound conclusion can be made on the
nature of the $\sigma$ resonance so far. Using large $N_c$
expansion technique and the analyticity property of scattering
amplitude, it was recently proved that though the $\sigma$
resonance has a large width when $N_c=3$, the pole does not remain
on the complex $s$ plane when $N_c$ increases.~\cite{XZ05}
Instead, it most likely moves to the real axis when
$N_c\to\infty$. What can not be excluded definitely is the
situation that the pole moves to $\infty$ when $N_c\to\infty$. The
uncertainty arises because it is not clear whether the $\sigma$
pole should dominate the low energy physics for arbitrary value of
$N_c$, though it indeed does when $N_c=3$.~\cite{zhou04} However
it seems to be a natural assumption that the sigma pole dominates
at low energies. In this note we will adopt such an assumption. We
will investigate the pole trajectory of the $\sigma$ resonance
when $N_c$ varies. The tools we use are [1,1] Pad\'e approximation
of $SU(3)$ chiral amplitude, [1,2] Pad\'e approximation of $SU(2)$
chiral amplitudes,  and the PKU parametrization form developed by
our group.~\cite{pku1,zhou04}
 Part of our results were already reported in
Ref.~\cite{pku2}. It should be noticed that all of these methods
are model dependent approaches. On one side, as will be made clear
in the following discussions, the role of one pole dominance
approximation is transparent in the PKU approach in order to trace
the pole trajectory. On the other side, the Pad\'e approximation
violates the \textit{very} important analyticity property of
scattering amplitudes,~\cite{Qin} even though it does not
explicitly rely on the additional sigma pole dominance assumption.
Analyticity property has been shown to be crucial in understanding
the $N_c$ properties of $S$ matrix poles.~\cite{XZ05} This point
will be further extinguished in the later discussions. Using [1,2]
Pad\'e method, we will construct an explicit counter-example  of
unitary $T$ matrix with correct $N_c$ counting rule, $T\sim
O(N_c^{-1})$. The $T$ matrix violates analyticity and the $\sigma$
pole in which resides on the complex $s$ plane when $N_c$ is
large. The $\sigma$ pole will finally annihilate with another pole
on the physical sheet, as dictated by the $N_c$ counting rule of
the $T$ matrix element. Nevertheless, such a phenomenon does not
happen in the realistic Pad\'e amplitudes and the predictions of
the latter on the $\sigma$ pole trajectory turn out to be very
similar to the output of the PKU approach. In the following we
will exhibit our discussions and results in different
approximations separately.

\section{$SU(3)$ [1,1] $\rm{Pad\acute{e}}$ approximation}\label{sect3}

\begin{figure}
\begin{center} \mbox{\epsfxsize=5cm \epsffile{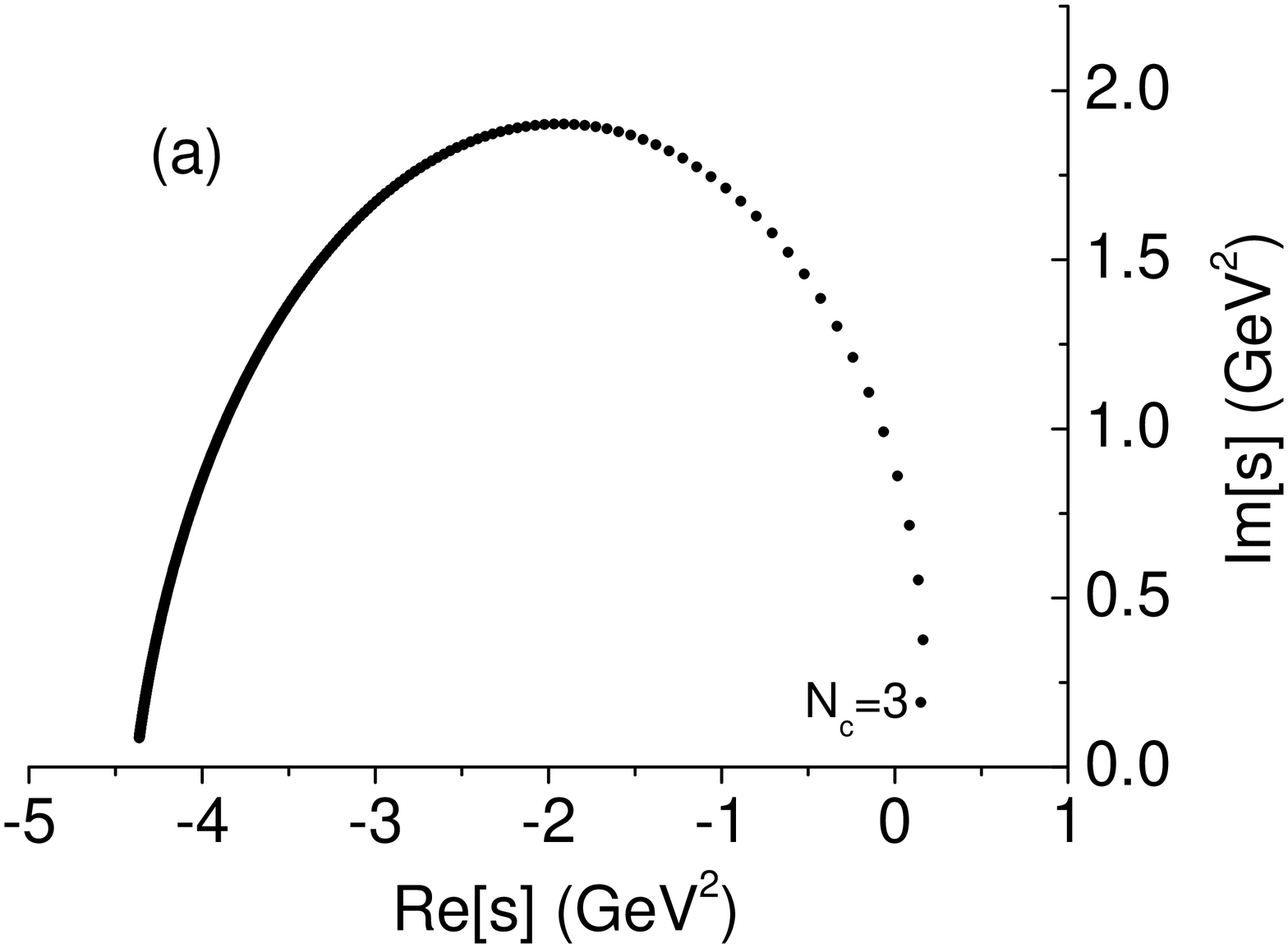}} \hspace{1cm}
\mbox{\epsfxsize=5cm \epsffile{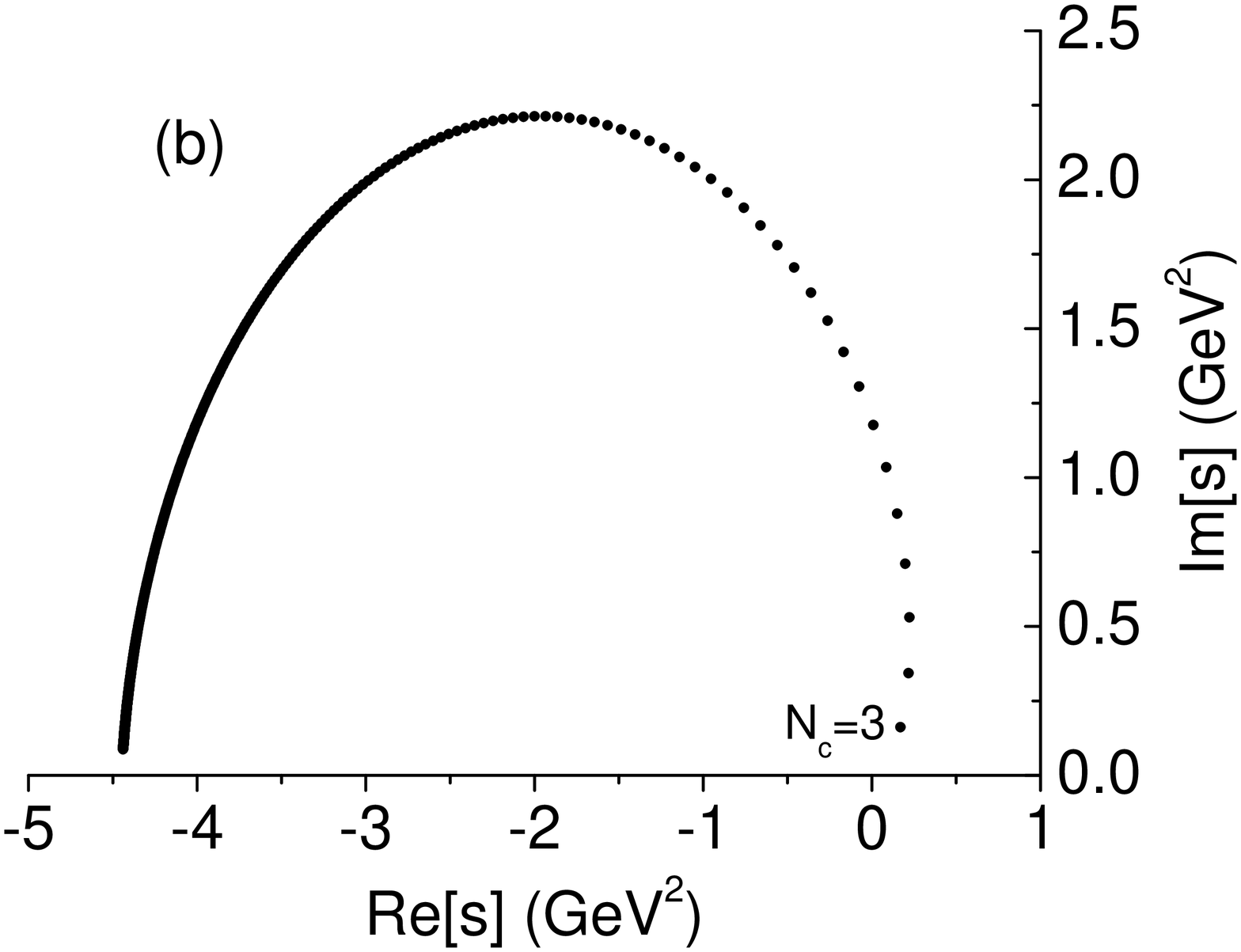}}
\end{center}\caption{\label{ppade}
$\sigma$ pole's trajectory as $N_c$ increase on  complex $s$ plane
with [1,1] $\rm{Pad\acute{e}}$ approximation method: a) $SU(3)$;
b) $SU(2)$.}
\end{figure}

The [1,1] $\rm{Pad\acute{e}}$ amplitude can be written as,
\be\label{amp}
T^{I[1,1]}_{J}(s)=\frac{T^{I}_{J,2}(s)}{1-{T^{I}_{J,4}(s)}/{T^{I}_{J,2}(s)}}\ee
where $T^{I}_{J,n}(s)$ is the partial wave amplitude of O($p^{n}$)
obtained from SU(3) chiral perturbation theory
(ChPT).~\cite{BKM91} The corresponding $S$ matrix is,
 \be
S^{[1,1]}(s)=1+2i\rho(s)T^{[1,1]}(s)\ . \ee
 Then we can use numerical
method to trace its poles' behavior when varying the variable
$N_c$.
 We illustrate
one typical trace on the complex $s$ plane in Fig.~\ref{ppade}a
with the values of low energy constants chosen in
Table.~\ref{tab1}. From it we can see that when $N_c$ is large
enough the $\sigma$ pole will move towards negative real $s$ axis
on the complex $s$ plane. However, it is not a unavoidable fate
for the $\sigma$ pole. The $\sigma$ pole's destination depends
directly on the value of parameters $L_2$ and $L_3$.
\begin{table}
\begin{center} \caption{\label{tab1}$SU(3)$ low energy
constants{\protect\cite{Pelaez2}} and their leading $N_c$ order. }
{\begin{tabular}{|c|c|c|}\hline
 Parameter& $\rm{Pad\acute{e}}$ 
&  $N_c$ leading\\
$\times10^{-3}$&$\mu=770Mev$&  behavior\\
 \hline
$2L_{1}-L_{2}$  & $ 0.0\pm0.1 $ &   $O(1)$\\
$L_{2}$  & $ 1.18\pm0.1 $ &  $O(N_{c})$\\
$L_{3}$  & $ -2.93\pm0.14 $ & $O(N_{c})$\\
$L_{4}$  & $ 0.2\pm0.004 $ &   $O(1)$\\
$L_{5}$  & $ 1.8\pm0.08 $ &  $O(N_{c})$\\
$L_{6}$  & $ 0.0\pm0.5 $ &   $O(1)$\\
$L_{7}$  & $ -0.12\pm0.16 $ &  $O(1)$\\
$L_{8}$  & $ 0.78\pm0.7 $ &  $O(N_{c})$\\
\hline
\end{tabular}}
\end{center}
\end{table}
This is understood since actually the $\sigma$ pole location in
[1,1] Pad\'e amplitude when $N_c$ is large can be solved
analytically in the chiral limit:
 \bqa
 \label{solu1}
s_{\sigma}&=&\frac{3F_{\pi}^{2}}{44L^{}_{1}+28L^{}_{2}+22L^{}_{3}}\
, \eqa
 where $F_{\pi}$ is the $\pi$ decay constant and
$L^{}_{i}$ are the  low energy constants of the chiral lagrangian.
It is verified numerically the explicit chiral symmetry breaking
term does not disturb the $\sigma$ pole trajectory much. From the
above solution we can see that $\sigma$ pole's position in large
$N_c$ limit is sensitive to the coupling constants $L_{1}$, $
L_{3}$ in the large $N_c$ limit where $2L_{1}=L_{2}$\cite{GL85}.
We observe the following behavior for the destination of $\sigma$
 \bqa
50L_{2}+22L_{3}\left\{\begin{array} {ll}
 >0, & \mbox{ $\sigma$ approaches positive real s axis; } \\ =0, & \mbox{
  $\sigma$ moves to $\infty$; } \\<0, & \mbox{ $\sigma$ approaches negative s axis; }
\end{array} \right.
 \eqa
Especially we have numerically checked the case when
$50L_{2}+22L_{3}=0$ and the result confirms the above conclusion.
Except such a peculiar choice of parameters the pole will always
move towards the real $s$ axis in the large $N_c$ limit. It can
also be proved that except the particular case the  $\sigma$
pole's width $\sim O(1/N_c)$. Thus according to Ref.~\cite{Lebed},
the $\sigma$ meson behaves like a normal $q\bar q$ in the large
$N_c$ limit. When $N_c$ is finite, from Fig.~1 we realize that it
is heavily renormalized by the $\pi\pi$ continuum.

\section{ $SU(2)$ [1,1] and [1,2]$\rm{Pad\acute{e}}$ approximation}\label{sect4}

We also studied $00$ channel using the $SU(2)$ [1,1] Pad\'e
amplitude. By solving the $N_c$ leading order $S$ matrix's pole
position and using the relation between coupling constants
$l_{1},...,l_{4}$ at
$O(p^{4})$ and $SU(3)$ coupling constants $L_{i}$,
we find in the chiral limit the solutions in Eq.~(\ref{solu1})
appear again. Also the pole trajectory appears to be similar
comparing with the SU(3) case, see fig.~1b).

There are new coupling constants enter the $O(p^{6})$ scattering
amplitude\cite{Bijnens1} listed in the table~$2$ together with
their $N_c$ dependence. The partial wave unitary amplitude of
$[1,2]$ $\rm{Pad\acute{e}}$ approximation is,
 \bqa
T^{I[1,2]}_{J}(s)=\frac{T^{I}_{J,2}(s)}{1-\frac{T^{I}_{J,4}(s)}
{T^{I}_{J,2}(s)}-\frac{T^{I}_{J,6}(s)}{T^{I}_{J,2}(s)}+(\frac{T^{I}_{J,4}(s)}{T^{I}_{J,2}(s)})^{2}}
\eqa

\begin{figure}
\begin{center} \mbox{\epsfxsize=5cm \epsffile{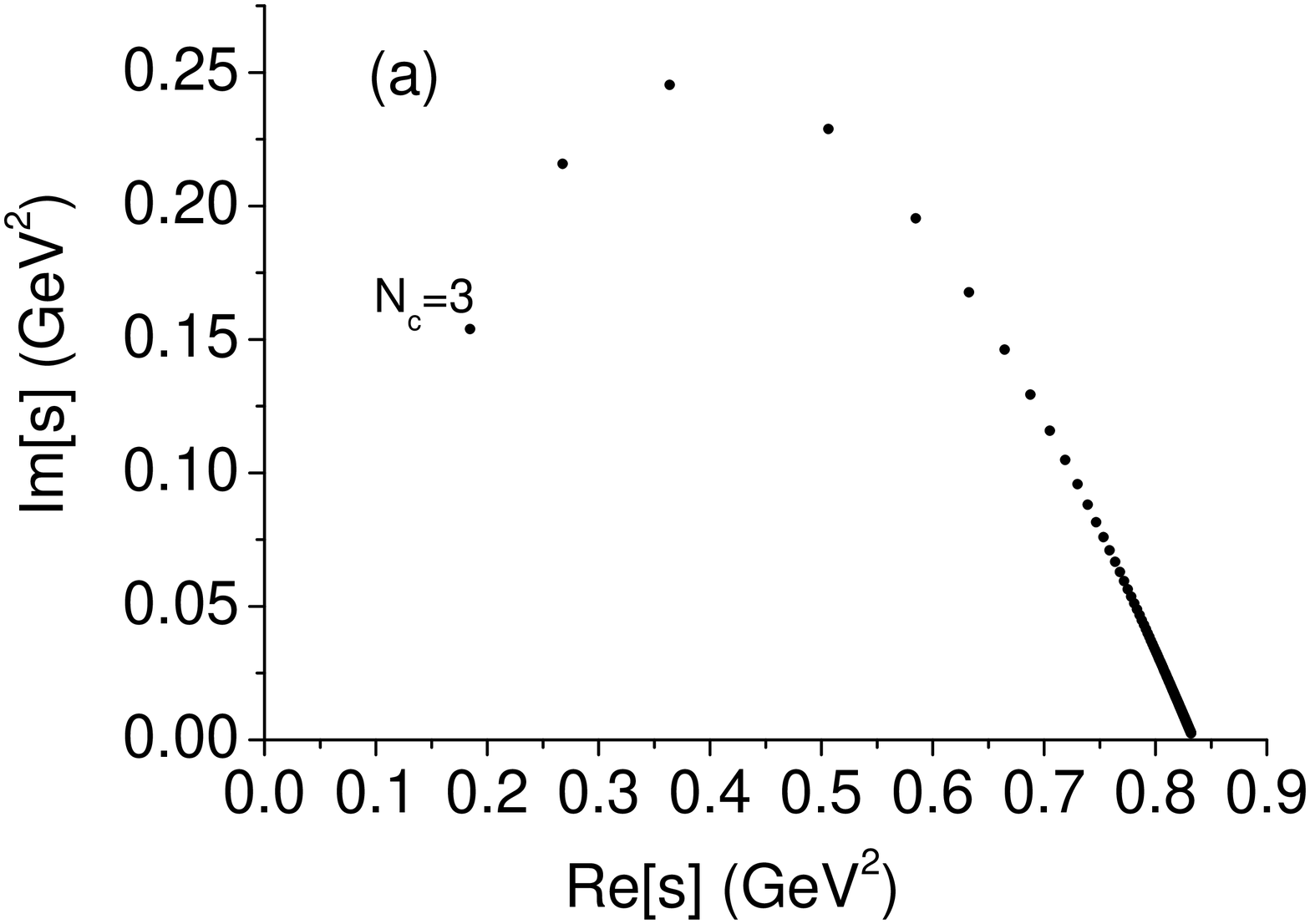}} \hspace{1cm}
\mbox{\epsfxsize=5cm \epsffile{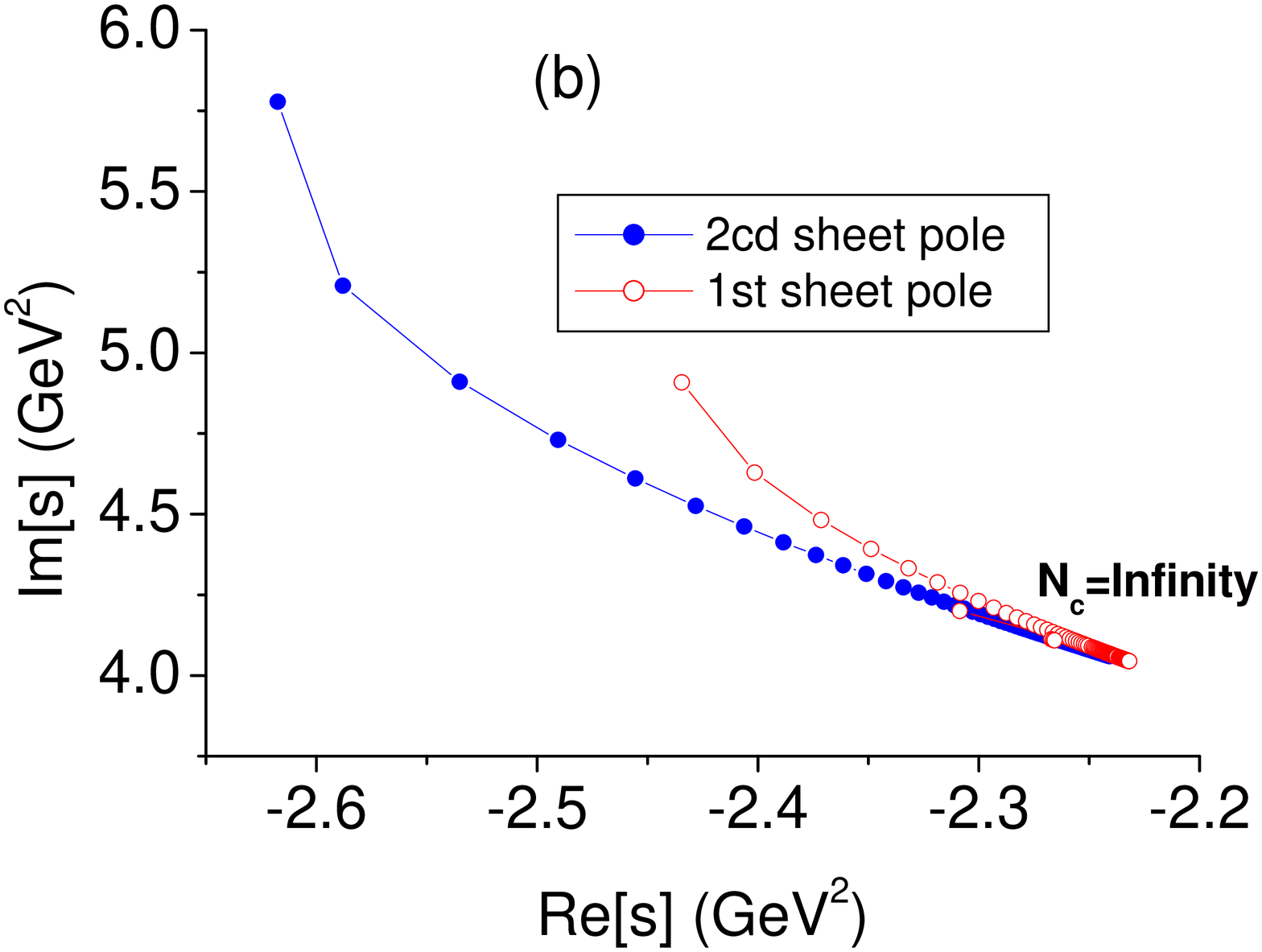}}
\end{center}\vspace*{8pt}\caption{\label{su212}
a) The $\sigma$ pole trajectory on $s$ plane from [1,2]
$\rm{Pad\acute{e}}$ amplitude; b) poles on different sheets cancel
each other in large $N_c$ limit. The starting points of the two
trajectories correspond to arbitrary values of $N_c$ rather than
3. }
\end{figure}

The $T^{I[1,2]}_{J}(s)$ amplitude predicts some additional
spurious poles of the $S$ matrix\cite{Qin} and to solve the
leading $N_c$ order $S$ matrix pole position analytically is
unavailable. By using numerical method, it is found that the
$\sigma$ pole also moves to the real $s$ axis on the complex $s$
plane in the large $N_c$ limit. We illustrate one of the typical
$\sigma$'s trace  in Fig.~\ref{su212}a) using the result of
table~2. It is worth noticing that in the SU(2) [1,2] Pad\'e
amplitude, according to our extensive but non-exhausting numerical
search, the $\sigma$ pole always moves to the positive real axis,
rather than the negative real axis, when $N_c\to\infty$.
\begin{table}
\centering \caption{\label{tab2}$SU(2)$ ChPT coupling constants
and their leading $N_c$ scaling{\protect\cite{Bijnens2}}}
{\begin{tabular} {|c|c|c|} \hline
Parameter & $ \rm{Pad\acute{e}}$   & $N_c$ leading\\
$ $& $\mu=770MeV$ & behavior\\
 \hline
$10^{3}l_{1}^{r}$  & $ 3.4 $ &  $O(N_{c})$\\
$10^{3}l_{2}^{r}$  & $ 4.8 $ &  $O(N_{c})$\\
$10^{3}l_{3}^{r}$  & $ 0.7 $ &  $O(N_{c})$\\
$10^{3}l_{4}^{r}$  & $ 9.4 $ &  $O(N_{c})$\\
$10^{4}r_{1}^{R}$  & $ -0.6$ &  $O(N_{c}^{2})$\\
$10^{4}r_{2}^{R}$  & $ 1.3 $ &  $O(N_{c}^{2})$\\
$10^{4}r_{3}^{R}$  & $ -1.7$ &  $O(N_{c}^{2})$\\
$10^{4}r_{4}^{R}$  & $ -1.0$ &  $O(N_{c}^{2})$\\
$10^{4}r_{5}^{R}$  & $ 1.1 $ &  $O(N_{c}^{2})$\\
$10^{4}r_{6}^{R}$  & $ 0.3 $ &  $O(N_{c}^{2})$\\
\hline
\end{tabular}}
\end{table}

Our numerical analyses indicate that the $\sigma$ pole's $N_c$
trajectory follows the rules given in Ref.~\cite{XZ05}. But since
analyticity property is used in deriving the conclusions made in
Ref.~\cite{XZ05}, there could exist exceptional situation if
analyticity is not obeyed. Indeed we find such a pedagogical
example: If those coupling constants $r_{i}$ are set to scale as
$N_c$, though it's not physical, the [1,2] Pad\'e amplitude is
still unitary and O($1/N_c$). In this situation, we find that in
the large $N_c$ limit the `$\sigma$' pole doesn't move to the real
$s$ axis any more. It just stays on the complex $s$ plane in the
large $N_c$ limit. It is found that there is another  pole in the
physical sheet which approaches to the `$\sigma$' pole and they
annihilate with each other when $N_c=\infty$. We have plotted such
an example in Fig.~\ref{su212}b). From the discussions made in
Ref.~\cite{XZ05}, it is realized that the annihilation phenomenon
has to happen as dictated by the fact that $T\sim O(N_c^{-1})$.

\section{The PKU approach}\label{sect5}
In this section we study the pole trajectory in the PKU
parametrization\cite{pku2} form in the $IJ=00$ channel. The
advantage of using the PKU parametrization form comparing with the
Pad\'e approximation  is that it avoids the problem of the
violation of analyticity by the latter.\cite{notice}
 We firstly expand $T$ matrix from SU(2) $O(p^4)$ ChPT amplitude at $2\pi$ threshold,
 \bqa\label{AAA}
T^{ChPT}(s)=t_{0}+i
t_{1}(\frac{s-4m_{\pi}^{2}}{m_{\pi}^2})^{\frac{1}{2}}
+t_{2}(\frac{s-4m_{\pi}^{2}}{m_{\pi}^{2}})+i
t_{3}(\frac{s-4m_{\pi}^{2}}{m_{\pi}^{2}})^{\frac{3}{2}}+
\cdot\cdot\cdot\eqa
 where those coefficients $t_{i}$ are functions of the low energy constants.
 On the other side the $T$ matrix in the PKU parametrization form is parametrized
 as,\cite{pku2}
  \be
T(s)=(\frac{-s+M^{2}+i\rho(s)sG}{-s+M^{2}-i\rho(s)sG}e^{2i\rho(s)f(s)}-1)/(2i
\rho(s))
 \ee
where we have assumed one pole (and hence the $\sigma$ pole)
dominance. To expand above parameterized $T$ matrix at
$4m_{\pi}^{2}$:
 \bqa\label{BBB} T(s)=
C_{0}+i\frac{C_{1}}{m_{\pi}}(s-4m_{\pi}^{2})^{1/2}
+\frac{C_{2}}{m_{\pi}^{2}}(s-4m_{\pi}^{2})
+O((s-4m_{\pi}^{2})^{3/2})\ , \eqa
 The background contribution $f$ is
obtained from ChPT and the threshold expansion,
 \be
f(s)=f_0+\frac{f_1}{m_\pi^2}(s-4m_\pi^2)+\frac{f_2}{m_\pi^4}
 (s-4m_\pi^2)^2+O((s-4m_\pi^2)^3)\ .
 \ee
Matching Eq.~(\ref{AAA}) and Eq.~(\ref{BBB}) at each order
leads to for example the following equations:
 \be\label{match}
C_0=t_0\ ,\,\,\, C_2=t_2\ ,\,\,\, C_4=t_4\ .
 \ee
There are only two parameters (for one pole) with 3 equations and
the system is over determined. Therefore 3 different solutions of
the $\sigma$ pole exist depending on which two out of 3 equations
we choose. Nevertheless it is found that the three solutions are
highly compatible for realistic situation when
 $N_c=3$, the approximate solution is roughly
$M_{\sigma} \simeq 460 MeV$, $\Gamma_{\sigma} \simeq 470
 MeV $, which agrees rather well with more rigorous
 calculations\cite{CGL01,zhou04}. In fact, one can prove that the
 3 solutions become unique in the large $N_c$ and chiral limit.
Solving $C_0=t_0$ one gets \be \label{GGG}
G=-\frac{1}{4m_\pi^2}(-4m_\pi^2+{M^2})({f_0}-{t_0})\ .
 \ee
  We can
get $M$ by solving $C_2=t_2$,
 \bqa \label{MMM}
M=\frac{2m_\pi\,{\sqrt{{{f_0}}^3 + 12\,{f_1} - 3\,{{f_0}}^2\,{t_0}
+ 3\,{f_0}\,{{t_0}}^2 -
        3\,{{t_0}}^3 - 12\,{t_2}}}}{{\sqrt{-3\,{f_0} + {{f_0}}^3 + 12\,{f_1} + 3\,{t_0} -
       3\,{{f_0}}^2\,{t_0} + 3\,{f_0}\,{{t_0}}^2 - 3\,{{t_0}}^3 - 12\,{t_2}}}}\eqa
In the large $N_c$ limit the above solution can be simplified as,
 \bqa \label{LNCGM1} G&=&-\frac{1}{4m_\pi^2}(-4m_\pi^2+{M^2})(f_0-{t_0})
        \nonumber \\M&=&\frac{4m_\pi\,{\sqrt{{t_2}}}}
  {{\sqrt{f_0-{t_0} + 4\,{t_2}}}}\eqa
Where up to $O(p^4)$, $t_0={7m_\pi^2\over {{32\pi F_{\pi
}}}^2}+\frac{m_\pi^4}{2{{{{{\pi F}}_{\pi
}}}^4}}(15L_2+5L_3+L_5+5L_8)$ , $t_2={m_\pi^2\over {{16\pi F_{\pi
}}}^2}
  +{m_\pi^4\over 2
   {{{{\pi F}_{\pi }}}^4}}(10L_2+4L_3+L_5)$,  $f_0=-|T^{\chi PT}(0)|=-\frac{m_\pi^2}
    {32{{{{\pi F}_
           {\pi }}}^2}}+
   \frac{m_\pi^4}
    {6{{{{\pi F}_{\pi }}}^
        4}}(25L_2+11L_3-9L_5+15L_8)$.
  It is verified that the above solution satisfies the equation
$C_4=t_4$ in the large $N_c$ and chiral limit and agrees exactly
with Eq.(\ref{solu1}).
  From Eq.(\ref{MMM}), we see that $M$ is $O(1)$ and $G$ is
  $O(1/N_c)$.\cite{Gamma}

%

The pole trajectory with respect to the variation of $N_c$ can be
traced numerically. The result is found to be very similar to
[1,1] $\rm{Pad\acute{e}}$
  result. The plot of the pole trajectory can be found in
  Ref.~\cite{pku2}

\section{Summary}\label{sect6}

Using Pad\'e approximation and the unitarization approach
developed by our group, we discuss the $N_c$ dependence of the
mass and width of the $\sigma$ resonance. This paper is a
supplementary to Ref.~\cite{XZ05}. Though
 the very existence of
the $\sigma$ pole has been firmly established, it's dynamical
nature remains mysterious. We hope the discussion given here will
be of some use towards a deeper understanding on its origin. Our
investigation seems to support the picture that the $\sigma$ pole
is a conventional $\bar qq$ meson environed by heavy pion clouds.
However the possibility that the $\sigma$ pole moves to the
negative real axis as revealed by our investigation makes the
topic interesting, which certainly deserves further
investigations. Another scenario which can not be excluded is that
the $\sigma$ pole moves to $\infty$, this situation may happen
when the $\sigma$ pole does not contribute to the $L_i$ parameters
of the low energy effective theory. We emphasize here that we
disagree with  the suggestion that the $\sigma$ pole is a $\bar
q\bar qqq$ state. In fact, under mild assumptions, it is
demonstrated that no resonance state with $M\sim O(1)$,
$\Gamma\sim O(1)$ could exist, and therefore  a tetra quark state
does not exist in general.\cite{XZ05,menu04} This result is a step
forward to the previous understanding on such issue.\cite{Lebed}


\begin{thebibliography}{0}
\bibitem{notice}Notice that poles does not automatically appear   on the physical sheet in here.
It is the solutions of Eq.~(\ref{match}) tells on which sheet
poles locate.
\bibitem{Gamma}Thus does not go  to $\infty$. But this is a consequence of
the assumption that the $\sigma$ pole doiminates low energy
physics at large $N_c$.
\end{thebibliography}
\end{document}